# Effects of Disorder on the Pressure-Induced Mott Transition in $\kappa$-(BEDT-TTF)$_2$Cu[N(CN)$_2$]Cl

**Elena Gati [1,*], Ulrich Tutsch [1], Ammar Naji [1], Markus Garst [2], Sebastian Köhler [1], Harald Schubert [1], Takahiko Sasaki [3] and Michael Lang [1]**

1 Institute of Physics, SFB/TR49, Goethe University Frankfurt, Max-von-Laue-Straße 1, 60438 Frankfurt am Main, Germany; tutsch@physik.uni-frankfurt.de (U.T.); Naji@physik.uni-frankfurt.de (A.N.); s.koehler@physik.uni-frankfurt.de (S.K.); h.schubert@physik.uni-frankfurt.de (H.S.); michael.lang@physik.uni-frankfurt.de (M.L.)

2 Institute for Theoretical Physics, Technical University Dresden, Zellescher Weg 17, 01062 Dresden, Germany; markus.garst@tu-dresden.de

3 Institute for Materials Research, Tohoku University, Katahira 2-1-1, Sendai 980-8577, Japan; takahiko@imr.tohoku.ac.jp

* Correspondence: gati@physik.uni-frankfurt.de; Tel.: +49-69-798-47240

**Abstract:** We present a study of the influence of disorder on the Mott metal-insulator transition for the organic charge-transfer salt $\kappa$-(BEDT-TTF)$_2$Cu[N(CN)$_2$]Cl. To this end, disorder was introduced into the system in a controlled way by exposing the single crystals to X-ray irradiation. The crystals were then fine-tuned across the Mott transition by the application of continuously controllable He-gas pressure at low temperatures. Measurements of the thermal expansion and resistance show that the first-order character of the Mott transition prevails for low irradiation doses achieved by irradiation times up to 100 h. For these crystals with a moderate degree of disorder, we find a first-order transition line which ends in a second-order critical endpoint, akin to the pristine crystals. Compared to the latter, however, we observe a significant reduction of both, the critical pressure $p_c$ and the critical temperature $T_c$. This result is consistent with the theoretically-predicted formation of a soft Coulomb gap in the presence of strong correlations and small disorder. Furthermore, we demonstrate, similar to the observation for the pristine sample, that the Mott transition after 50 h of irradiation is accompanied by sizable lattice effects, the critical behavior of which can be well described by mean-field theory. Our results demonstrate that the character of the Mott transition remains essentially unchanged at a low disorder level. However, after an irradiation time of 150 h, no clear signatures of a discontinuous metal-insulator transition could be revealed anymore. These results suggest that, above a certain disorder level, the metal-insulator transition becomes a smeared first-order transition with some residual hysteresis.

## 1. Introduction

Organic charge-transfer salts of type $D_2A$, consisting of donor molecules $D$ and acceptor molecules $A$, have been established as model systems to explore the physics of strong electron correlations [1,2]. The role of Coulomb correlations is particularly enhanced in this material class due to weak intermolecular interactions, giving rise to a small bandwidth $W$, in combination with a low charge-carrier concentration and an electronic structure with reduced dimensionality. The latter two aspects render the screening of the long-range part of the Coulomb interaction less effective. A prominent manifestation of strong correlation effects is the Mott metal-insulator transition [3]. The notion is that the itinerant electrons in an approximately half-filled conduction band localize, once the mutual Coulomb repulsion for two electrons on the same site, $U$, exceeds the band width $W$. Although the above-sketched simple picture describes the basics of the Mott transition, the understanding of the rich phenomenology of correlated electrons close to the Mott transition continues to pose major challenges [4–6]. In particular, salts of the $\kappa$-(BEDT-TTF)$_2X$ family, with BEDT-TTF = bis(ethylenedithio)-tetrathiafulvalene and $X$ a monovalent anion, have attracted high attention as prime examples of a *bandwidth*-tuned Mott transition [7,8] at fixed band filling. In these materials, which are considered to exhibit an effectively half-filled conduction band due to the strong dimerization of the BEDT-TTF molecules, the bandwidth can be controlled via the application of pressure $p$. In this way, a transition from a Mott insulating to a metallic state can be induced with increasing pressure through the accompanying decrease in the correlation strength $U/W$. Notably, already a weak pressure of $p \approx 30$ MPa is sufficient to induce the Mott metal-insulator transition in the system $\kappa$-(BEDT-TTF)$_2$Cu[N(CN)$_2$]Cl (abbreviated as $\kappa$-Cl hereafter) at low temperatures $T < 35\,\mathrm{K}$ [9,10]. This $p$- and $T$-range is very convenient for experiments aiming at fine tuning a material around the Mott transition under well-controlled conditions. Our present understanding of the Mott transition has benefited significantly from a number of detailed experimental studies on $\kappa$-Cl [11–14] and some transition-metal oxides, see, e.g., [15,16], together with advances from theoretical side through dynamical mean-field theory calculations [17]. It is by now well established that the Mott transition is of first order which terminates in a second-order critical endpoint at $(T_c, p_c)$ as there is no symmetry breaking accompanying the transition. However, the critical behavior and the underlying universality class have been a matter of debate. These aspects have been discussed intensively for $\kappa$-Cl [18–25], where the second-order critical end-point is located at $(T_c, p_c) \approx (36.5\,\mathrm{K}, 23.4\,\mathrm{MPa})$. Recently, based on a careful study of the lattice effects, it has been argued in Ref. [25] that the coupling of the correlated electrons to the lattice degrees of freedom is of crucial importance for understanding the Mott transition whenever the transition is amenable to pressure tuning. It has been found that this coupling alters the critical behavior from Ising criticality, for the purely-electronic Mott system, to mean-field criticality of an isostructural solid-solid endpoint in the presence of a compressible lattice, in accordance with the theoretical predictions [26,27]. It is worth noting that also the high-temperature behavior at $T \gg T_c$ has attracted considerable interest recently, as it shows characteristic signatures of a hidden quantum-critical point [28–30]. It has been suggested that quantum criticality is responsible for the unusual transport properties in the incoherent regime close to the Mott transition.

Another aspect of ongoing interest relates to the effects of disorder in strongly correlated electron systems. Actually, the questions surrounding the interplay of disorder and correlation [31–33] trace back to the different mechanisms for metal-insulator transitions in solids: Whereas the Mott transition is driven by electronic correlations, backscattering of non-interacting electronic waves from disorder, i.e., randomly distributed impurities, can induce the so-called *Anderson* insulating state [34,35]. In a naive picture, one would assume that the simultaneous action of disorder and correlations stabilizes the insulating phase. However, it has been found that disorder and correlations combine in a highly non-trivial way and can compete with each other [31,33]. Current research [36–41] addresses the following aspects: Do characteristic properties of the Mott insulator survive when introducing disorder in the strongly correlated system? Or, alternatively, how stable is the Anderson insulator when the correlations in such a disordered system become strong? Understanding this subtle interplay at the so-called Mott-Anderson-transition is of fundamental importance for a more realistic modeling of the behavior of real materials where disorder is inevitable. For advancing our understanding in this field, materials which allow for a control of both the strength of correlations as well as the level of disorder are required. Both aspects are met favorably in the family of organic charge-transfer salts.

Approaches to intentionally introduce disorder in a well-controlled manner in the $\kappa$-(BEDT-TTF)$_2X$ salts are manifold [42–51]. Among these, X-ray irradiation with typical irradiation times $0\,\mathrm{h} \leq t_{irr} \leq 600\,\mathrm{h}$ has been widely used to control disorder in a non-reversible manner (see Ref. [48] for a review). It was demonstrated that X-ray irradiation mainly creates molecular defects in the anion layer, whereas it leaves the conducting BEDT-TTF layer intact. The defects in the anion layer thereby create a random potential for the hole carriers in the BEDT-TTF layer. The power of this technique was demonstrated for $\kappa$-(BEDT-TTF)$_2$Cu[N(CN)$_2$]Br in which a transition from a correlated metal to an Anderson insulator [50] accompanied by a suppression of superconductivity [49] was observed. In addition, long-range antiferromagnetic order [52] in the Mott insulator $\kappa$-Cl, which is understood as an effect of strong $U$, was suppressed [53,54] by long irradiation $t_{irr} = 500$ h. It was argued that these results [53] highlight the role of disorder in forming a spin liquid close to the Mott transition.

Here, we focus on yet another aspect of the Mott transition in the presence of disorder by asking the questions: (i) How does quenched disorder influence the location of the Mott transition, i.e., the first-order transition line and its second-order critical endpoint in the $T - p$ phase diagram; and (ii) Is there any influence of disorder on the Mott criticality? To this end, we present a study of the combined effects of X-ray-induced disorder and pressure on the Mott insulator $\kappa$-Cl by means of thermodynamic and transport probes. Importantly, the use of He-gas pressure allows us to fine-tune the system across the Mott transition so that the *critical* regime in the phase diagram can be accessed experimentally.

The paper is organized as follows: In Sections 3.1–3.3, we present a detailed comparison of thermal expansion data around the Mott transition of a $\kappa$-Cl crystal in its *pristine* (non-irradiated) form with those of a crystal from the same batch after an exposure to X-ray irradiation for 50 h. For tracing the Mott transition and its critical endpoint in the $T - p$ phase diagram to higher disorder levels, i.e., irradiation doses up to 150 h, we present in Sections 3.4 and 3.5 the results of electrical resistance measurements on one single crystal at various disorder levels obtained after 50 h, 100 h and 150 h irradiation time. The results will be discussed in the context of available theoretical results in Section 4.

## 2. Materials and Methods

Single crystals of $\kappa$-(BEDT-TTF)$_2$Cu[N(CN)$_2$]Cl were grown by electrochemical crystallization following the standard procedure [55]. For the present study, crystals from two different batches #AF063 and #5-7 were used. The dimensions of the crystals are typically $0.3 \times 0.6 \times 0.7\,\text{mm}^3$ for #AF063 and $0.5 \times 0.3 \times 0.3\,\text{mm}^3$ for #5-7, respectively.

Samples were X-ray irradiated [48] at 300 K by using a non-filtered tungsten tube with 40 kV and 20 mA. The corresponding dose rate was estimated to be about 0.5 MGy/h. During the irradiation, the resistance was monitored. The plate-like crystals were irradiated from both sides. The irradiation time $t_{irr}$ denotes the overall exposure time.

Thermal expansion and resistance measurements were performed under variable He-gas pressure $p$. To this end, a copper beryllium (CuBe) pressure cell was used which was connected to a large-volume (50 L) helium gas bottle for thermal expansion measurements at $p \leq 30$ MPa or a helium-gas compressor for resistance measurements at $p \leq 100$ MPa. This technique enables measurements to be performed as a function of $T$ at constant $p$. In combination with the use of a He-gas bottle, the thermal expansion can also be measured as a function of $p$ at constant $T$. Pressure was determined by a pressure sensor (InSb) in the pressure cell and a manometer situated close to the high-temperature reservoir.

Thermal expansion measurements were performed using an ultra-high resolution dilatometer with $\Delta L/L \geq 5 \times 10^{-10}$. The setup was described in detail in Ref. [56]. The data analysis was performed in an identical procedure as described in Ref. [25].

The resistance was measured by an AC resistance bridge (LR 700, Linear Research, San Diego, CA, USA) in a standard four-terminal configuration. The current was applied perpendicular to the conducting layer. Contacts were made by graphite paste and 20 μm gold wires on which the sample was suspended. The temperature was controlled by a LakeShore 340 Temperature Controller (Lake Shore Cryotronics Inc., Westerville, OH, USA). Slow warming and cooling rates of ±0.2 K/min were used to ensure thermal equilibrium at all $T$ and $p$.

## 3. Results

### 3.1. Effects of Irradiation on the Lattice Effects at the Mott Transition in κ-Cl

Figure 1 shows data of the relative length change $\Delta L_b/L_b$, measured along the $b$ axis, as a function of pressure $p$ in the temperature range $30\,\mathrm{K} \leq T \leq 40\,\mathrm{K}$ for κ-Cl in its pristine form [25] (a) and after exposure to 50 h X-ray irradiation (b) (see Section 2 for a description of the methods used, including details of the irradiation procedure). The displayed data represent only the singular contribution to the relative length change which was obtained by subtraction of an in-$p$ linear background contribution from the $\Delta L_b(p)/L_b$ data. We start by recalling some basic notations from the study of the pristine sample in Ref. [25]. At the lowest temperature $T = 30\,\mathrm{K}$, a slightly broadened jump of the sample length as a function of pressure was observed at $p_{MI} \approx 22.5\,\mathrm{MPa}$ at which the length decreases on going from the low-pressure insulating side to the high-pressure metallic side of the transition. The observation of a discontinuity in $\Delta L_b/L_b$ is consistent with the fact that a first-order transition line is passed as a function of $p$. Upon increasing the temperature $T$, the discontinuity decreases in size and evolves into a continuous crossover behavior for $T \geq 37\,\mathrm{K}$. This data set discloses a strong coupling of the electronic degrees of freedom to the lattice close to the Mott transition. It manifests itself in strong non-linear variations of $\Delta L_b(p)/L_b$—a *breakdown of Hooke's law*—which can be observed in a wide range of *critical elasticity* of $\Delta T_c/T_c \approx 20\,\%$ above the critical endpoint. It was demonstrated in Ref. [25] that these lattice effects can be modeled by mean-field criticality. The important implication of this finding was that, although the Mott transition is driven by strong electron correlations, the coupling of the electrons to the lattice degrees of freedom eventually alters the Mott critical behavior to that of an isostructural solid-solid endpoint with mean-field criticality [26,27].

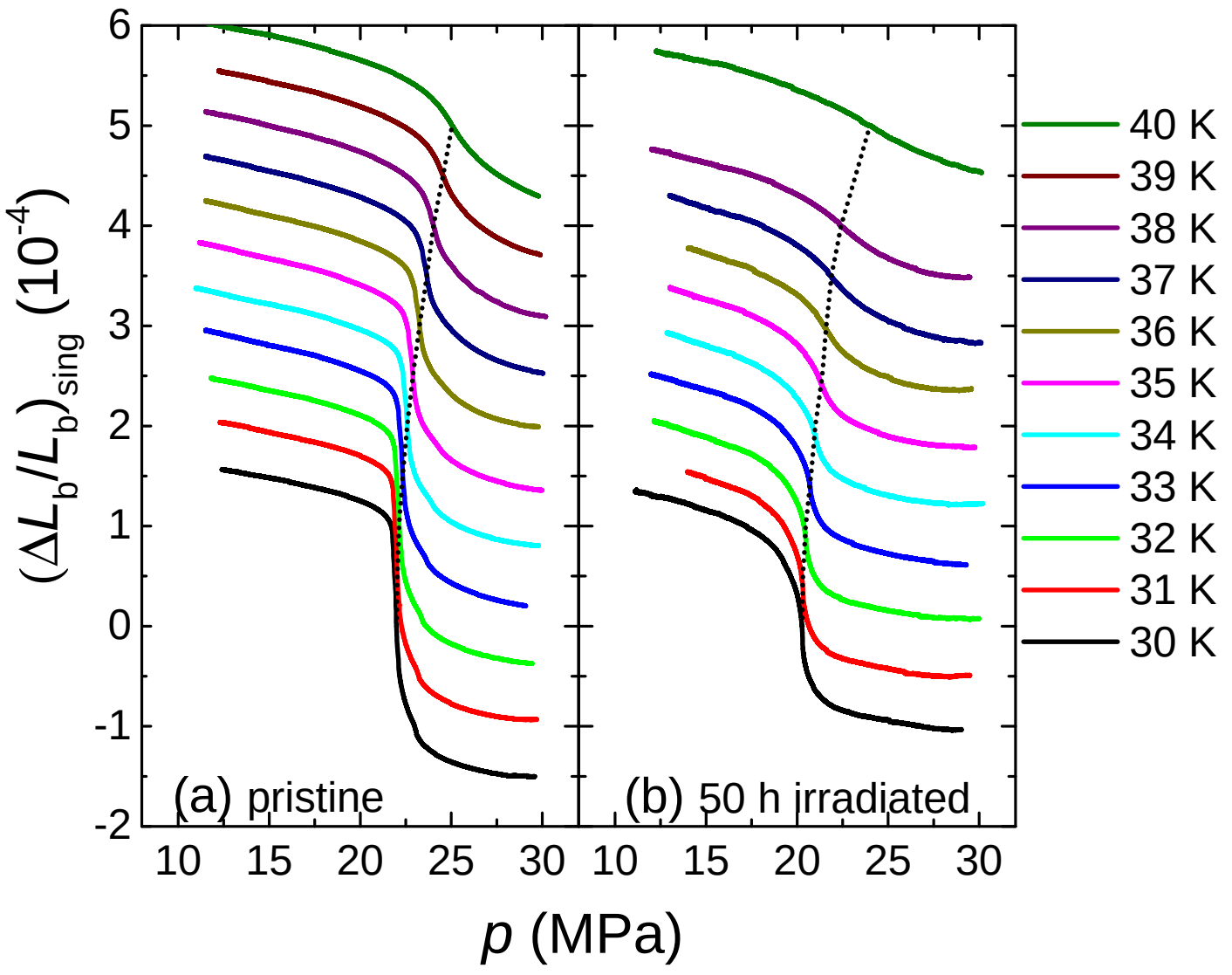

**Figure 1.** Singular part of the relative length change $(\Delta L_b/L_b)_{sing}$ of κ-(BEDT-TTF)$_2$Cu[N(CN)$_2$]Cl (batch #AF063) in its pristine form (data taken from Ref. [25]) (**a**) and after exposure to X-ray irradiation for 50 h (**b**). Data were taken along the out-of-plane $b$ axis as a function of pressure $p$ at constant temperatures between 30 K and 40 K.

A lattice response of similar strength around the Mott transition can also be inferred from the $\Delta L_b/L_b$ data set in Figure 1b taken on a crystal which was exposed to X-ray for 50 h. At the lowest temperature of $T = 30\,\mathrm{K}$, we find a strong decrease of the length upon going from the low-pressure insulating side to the high-pressure metallic side at $p_{MI} \approx 20.2\,\mathrm{MPa}$. Note that this $p_{MI}$ value is smaller by ≈1.8 MPa than the corresponding value for the pristine sample at 30 K. This shift of the transition to lower pressure upon increasing the degree of disorder is one of the central results of the present work and will be discussed below in more detail. Concerning the character of the transition for the irradiated crystal, we observe an even more broadened jump-like feature as compared to the pristine crystal where a small broadening is visible. Yet, the data of the irradiated crystal still reveal a discontinuous change of the lattice parameter, i.e., a first-order phase transition (see below for a detailed discussion of the determination of $T_c$ and $p_c$). In the pristine sample, the finite width of the transition is likely associated with a small amount of disorder that leads to a spatial variation of internal stress induced by impurities and/or other crystal defects. In order to account quantitatively for this broadening effect, we assumed in Ref. [25] a multi-domain state, caused by extrinsic defects, where each domain contributes independently to the mean strain. Experimentally, the broadening manifests itself in a peak of finite width in the compressibility $\kappa_b(p) = -\mathrm{d}(\Delta L_b/L_b)/\mathrm{d}p$, i.e., in the derivative of the present data sets with respect to $p$. In order to quantify the increase in broadening upon irradiation, we determine the width $\Delta p$ of the peak in $\kappa_b$ at $T = 30\,\mathrm{K}$ by its full width at half maximum. This procedure yields $\Delta p_{50h} = (0.81 \pm 0.05)\,\mathrm{MPa}$ for the irradiated sample and $\Delta p_0 = (0.38 \pm 0.01)\,\mathrm{MPa}$ for the pristine sample. This implies that the distribution of internal stress in the 50 h-irradiated sample is about twice as wide as in the pristine sample, giving rise to a larger disorder-affected pressure regime. By lacking a microscopic characterization of the defect structure in the irradiated samples, we can only speculate that the defect sites, created by irradiation in the anion layer, may cause an increased number of domains. We suspect that this also affects the internal stress of each domain which in turn leads to an overall wider stress distribution in such a crystal. As the technique of dilatometry is unable to reveal the required spatially-resolved information on the formation of domains, we omit a detailed discussion of the role of domain formation at the first-order transition here and treat this broadening on a phenomenological level, whenever necessary. In addition to an enhanced broadening, we find a significant reduction of the size of the discontinuity at $T = 30\,\mathrm{K}$ at $p_{MI}$ from $(\Delta L_b/L_b)|_{p_{MI}} = (2.3 \pm 0.1) \times 10^{-4}$ in the pristine case to $(\Delta L_b/L_b)|_{p_{MI}} = (1.5 \pm 0.1) \times 10^{-4}$ in the irradiated case. Apart from these small quantitative changes in the size and width of the anomalous length changes, we find that the evolution of the feature in $\Delta L_b/L_b$ to higher $T$ is qualitatively similar to the above-described pristine case: The size of the discontinuity becomes reduced with increasing $T$ until it is replaced by a continuous change of $\Delta L_b(p)/L_b$ for $T \geq T_c \approx 34\,\mathrm{K}$. Importantly, also for the irradiated sample, we find highly non-linear variations of the length as a function of pressure, even at temperatures way above $T_c$, see, e.g., the data at $T = 38\,\mathrm{K}$, where $\kappa_b(p = p_{MI})$, i.e., the slope of the $\Delta L_b/L_b$ curves at $p_{MI}$, exceeds $\kappa_b(p \ll p_{MI})$ by at least a factor of two. This behavior is distinctly different from the usual elastic behavior which is characterized by an in-first approximation linear change of $\Delta L_b/L_b$ vs. $p$, i.e., $\kappa \approx$ const., in accordance with Hooke's law of elasticity. The observed breakdown of Hooke's law here reflects the strong coupling between the critical electronic system and the underlying crystal lattice at the Mott transition, also in the presence of increased disorder.

### *3.2. T-p Phase Diagram for Weak Disorder*

In the following, we analyze the effects of disorder by focusing on certain characteristics of the Mott transition. This includes the precise location of the critical endpoint and the critical behavior in its surrounding. In order to determine the location of the critical endpoint ($T_c$, $p_c$) for the irradiated crystal with high accuracy from measurements of the relative length change $\Delta L_b/L_b$, we take two complementary approaches, similar to the procedure applied to the pristine sample in Ref. [25]. In our first approach, we determine ($T_c$, $p_c$) from the intersection of the first-order transition line with crossover lines which emanate from the critical endpoint. To this end, we combine results from $p$-dependent measurements, which allow a precise determination of the first-order transition line (see below), with those performed as a function of $T$. The latter measurements provide insight into the crossover lines by analyzing the maximum response $\alpha_{max}$ in the coefficient of thermal expansion $\alpha_b(T) = L_b^{-1}\mathrm{d}L_b/\mathrm{d}T$ for $T > T_c$, as proposed in Ref. [23]. Moreover, due to the finite slope of the first-order transition line at ($T_c$, $p_c$), these measurements as a function of temperature at different $p =$const. provide not only information on the crossover line for $T > T_c$ and $p > p_c$. In addition, these $T$-dependent experiments allow us to cross the first-order transition line for $T < T_c$ and $p < p_c$. As the crossing of the first-order transition is expected to manifest itself in a discontinuity in $\Delta L_b(T)/L_b$, corresponding to a divergent behavior in $\alpha_b(T)$, the observation of such discontinuities can serve as a criterion for localizing $T_c$. Figure 2 shows $\alpha_b(T)$, measured along the out-of-plane $b$ axis, of 50 h-irradiated $\kappa$-Cl at several constant $p$ values around $p_c$. Starting at low $p = 16.5$ MPa, we observe the usual increase of $\alpha_b$ with increasing $T$ without any signatures for anomalous behavior. As we approach the endpoint from the low-$p$ side, however, we observe sharp and pronounced anomalies in $\alpha_b(T)$ with the tendency to diverge in a narrow pressure range 20.3 MPa $\leq p \leq$ 21.0 MPa. We assign these features to signatures of the first-order phase transition. Note that for the data set taken at $p = 20.5$ MPa, the first-order transition line is crossed twice as a function of $T$ due to its $S$-shaped form. As a result, we find two successive sharp anomalies of opposite sign, the latter one corresponding to the reentrance from the intermediate-$T$ metallic state to the low-$T$ insulating state. In contrast, for $p > 21.0$ MPa, the divergent features in $\alpha_b(T)$ are absent and instead, $\alpha_b(T)$ shows a smooth anomaly at slightly higher $T$. These features can be assigned to signatures of a crossover line, as predicted theoretically in Ref. [23].

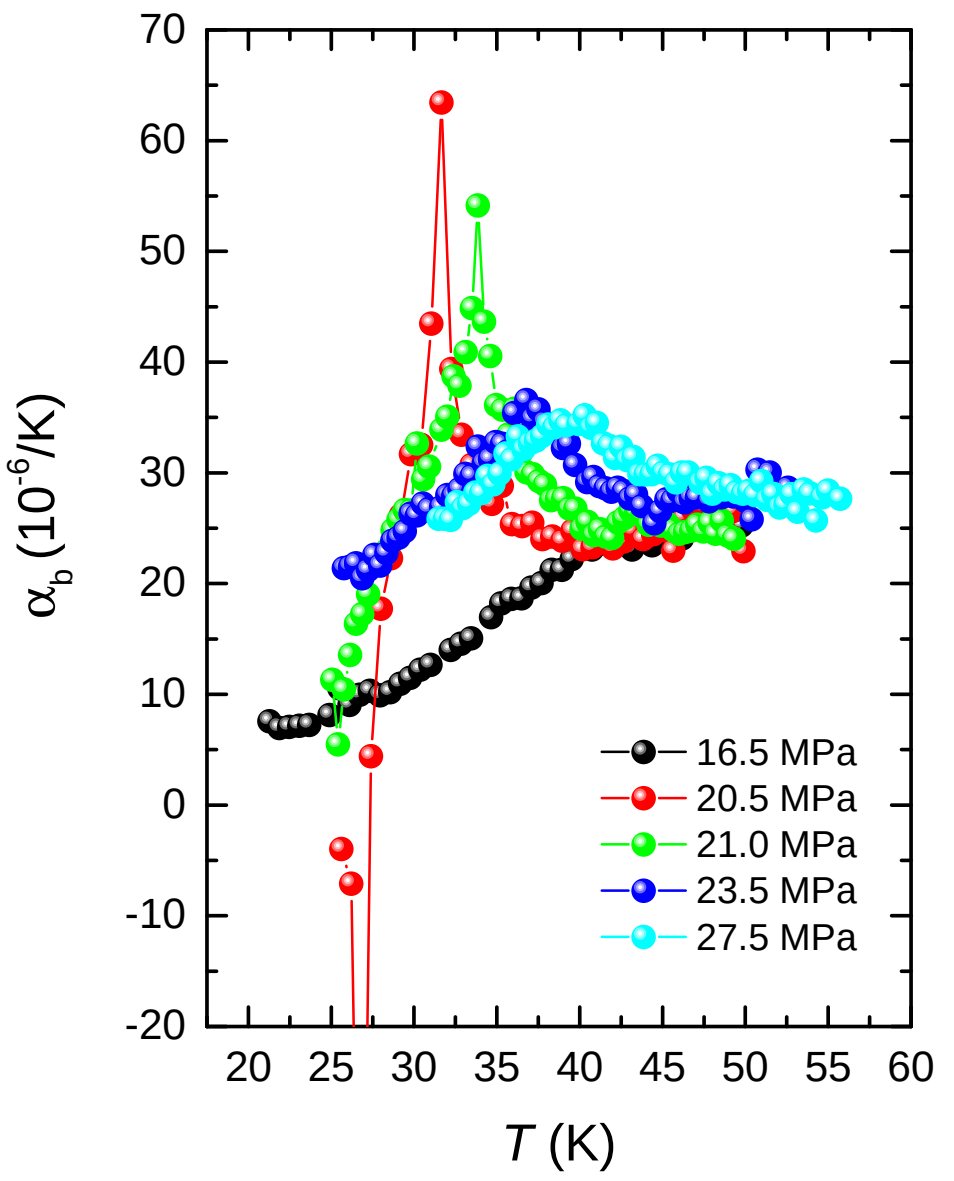

**Figure 2.** Thermal expansion coefficient along the out-of-plane $b$ axis, $\alpha_b$, of $\kappa$-(BEDT-TTF)$_2$Cu[N(CN)$_2$]Cl (batch #AF063) after exposure to X-ray for 50 h as a function of temperature $T$ at constant pressures 16.5 MPa $\leq p \leq$ 27.5 MPa.

In Figure 3, we compile in a $T$-$p$ phase diagram the position of the sharp peaks in $\alpha_b$ for $p \leq 21$ MPa, reflecting the first-order transition temperatures, and the positions of $\alpha_{max}$ for $p > 21.0$ MPa indicating crossover temperatures. In addition, we also include the positions of the inflection points in the $p$-dependent $\Delta L_b/L_b$ data sets at different constant $T$. These inflection points correspond to either the first-order transition for $T < T_c$ or to the so-called *Widom* line for $T > T_c$, i.e., the extension of the first-order transition line to higher $T$. From the intersection point of the first-order transition line determined from $p$-dependent measurements, and the crossover line determined from $T$-dependent measurements, we find $T_c = (34.0 \pm 0.5)$ K and $p_c = (21.0 \pm 0.2)$ MPa for the 50 h-irradiated sample.

An alternative approach to determine ($T_c$, $p_c$) is provided by analyzing the width of the features observed in $\Delta L_b(p)/L_b$. As mentioned above, the width of the discontinuity in $\Delta L_b/L_b$ for $T < T_c$ is governed by extrinsic effects. This gives rise to an in first approximation $T$-independent width for low $T$, as already visible in the bare data in Figure 1b. In contrast, for $T > T_c$ the width is strongly $T$-dependent and can therefore be assigned to the effects of criticality. For a quantification of the increase in the width upon increasing $T$, we determine the width of our experimental data by the full width at half maximum of the peak in $\kappa_b(p)$, as described above. We include the information of the width in the phase diagram, shown in Figure 3, by red-broken lines for the disorder-dominated regime at $T < T_c$ and blue-broken lines for the criticality-dominated regime at $T > T_c$. We find a significant increase in the width for $T > 36$ K. An extrapolation of the lines in the criticality-dominated regime from high $T \geq 36$ K, where disorder effects play only a minor role, to lower $T$ yields an alternative estimate of $T_c = (33.5 \pm 0.5)$ K. Note that the two different approaches yield an identical position of the critical endpoint within the error bars.

Now we can compare the so-derived phase diagram for the irradiated crystal with the one of the pristine sample. To this end, we include the position of the first-order line, the *Widom* line and the critical endpoint of the pristine sample in the phase diagram in Figure 3. We find that central properties of the Mott transition, i.e., the first-order character at low $T$ and the second-order critical endpoint, survive at the level of disorder introduced by 50 h of irradiation. However, this irradiation leads to a significant reduction of $T_c$ as well as $p_c$. Here, the exposure to X-ray for 50 h causes a decrease of $T_c$ by ≈2.5 K ($\Delta T_c/T_c \approx 7\,\%$) and of $p_c$ by ≈2.4 MPa ($\Delta p_c/p_c \approx 10\,\%$). This finding demonstrates a high sensitivity of the Mott transition line on the degree of disorder. We stress that the Mott transition lines for $\kappa$-Cl reported in the studies by Kagawa et al. [18] and Gati et al. [25], practically coincide, indicating a similar low level of disorder in the pristine crystals studied there. In addition, a study of pristine crystals from batch #AF063 (not shown) discloses that $T_c$ of two different crystals varies up to 1 K and $p_c$ up to 0.5 MPa. Thus, the variations in the disorder level of pristine samples are distinctly smaller than the one introduced by exposure to X-ray for 50 h.

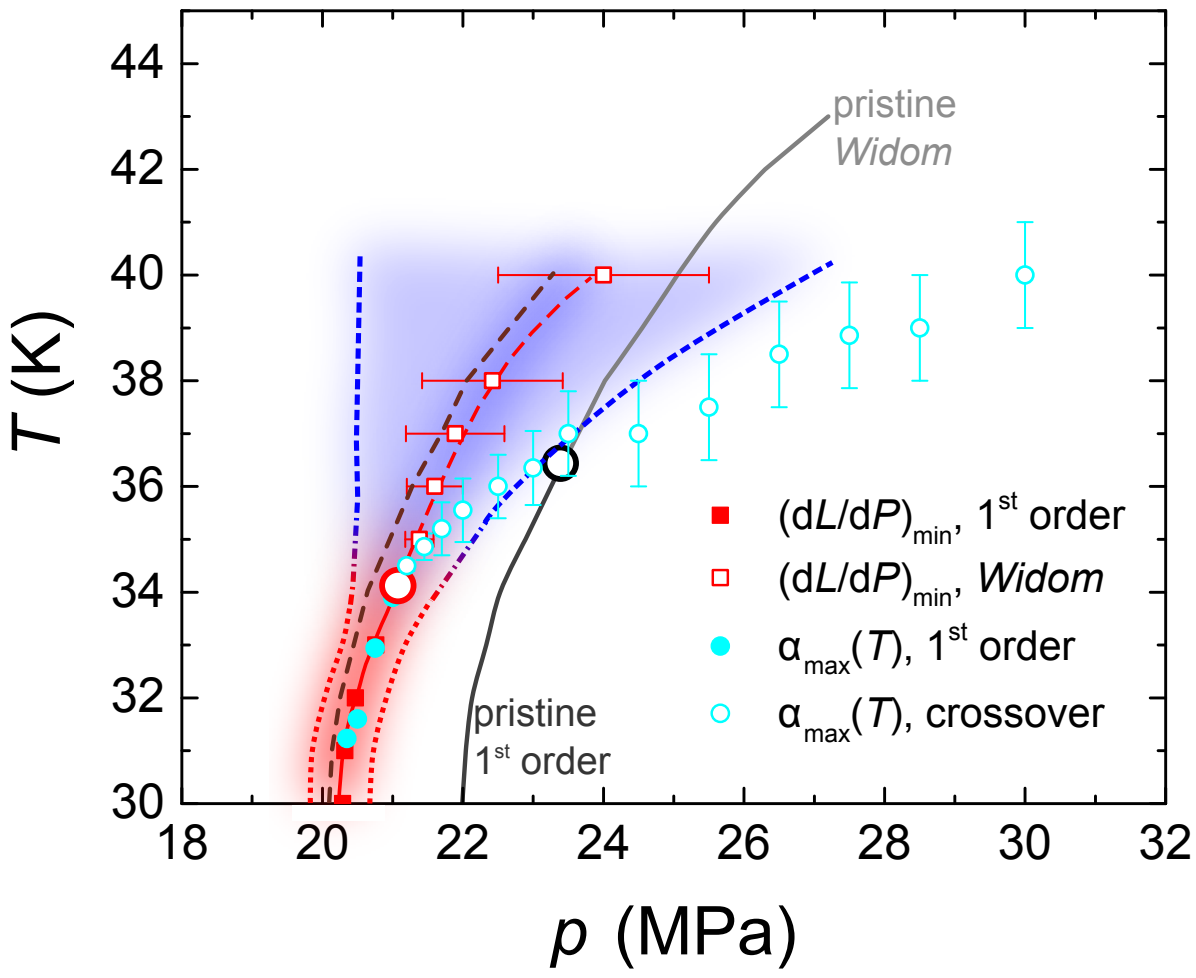

**Figure 3.** Experimentally determined temperature-pressure phase diagram of $\kappa$-(BEDT-TTF)$_2$Cu[N(CN)$_2$]Cl (batch #AF063) after exposure to X-ray for 50 h. Red full squares correspond to the first-order Mott transition line extracted from the inflection point of $\Delta L_b(p)/L_b$. Cyan full circles correspond to the first-order transition line extracted from temperature-dependent measurements. The latter data points were determined by the position of the maximum $\alpha_{max}$ of the thermal expansion coefficient $\alpha_b(T) = L_b^{-1}\,\mathrm{d}L_b/\mathrm{d}T$ for $p < p_c$. Red open symbols correspond to the *Widom* line. Open cyan symbols correspond to a crossover line, determined by the position of $\alpha_{max}$ for $p > p_c$. Red- and blue-shaded areas delimited by the broken lines in the same color code indicate the experimentally determined width of the features along the $b$ axis and can be assigned to the disorder-related (red) and the criticality-related (blue) crossover regimes, respectively. The broken lines represent, within the error margins, the full width at half maximum of the peaks in $\mathrm{d}L/\mathrm{d}P$. The brown dashed line corresponds to the $p_c(T)$ curve extracted from a fit of the data, presented in Figure 1 to the mean-field model of Equation (1) (see text for details). The first-order transition line (dark grey line), the *Widom* line (light grey line) and the critical endpoint (black open circle) for a pristine crystal of batch #AF063 [25] are given for comparison.

### 3.3. Critical Behavior for Weak Disorder

After having determined the location of the critical endpoint ($T_c$, $p_c$) for the irradiated crystal, we can proceed by analyzing the observed lattice effects in terms of the critical behavior. For pristine $\kappa$-Cl [25], the anomalous lattice effects could be well described by a mean-field critical model of an isostructural solid-solid endpoint. In fact, this type of criticality was predicted theoretically for the critical electronic Mott system that is coupled to a compressible lattice as a consequence of the long-ranged shear forces of the lattice [26,27]. In light of the strong lattice response revealed for the irradiated system, we employ the same mean-field critical model to the present $\Delta L_b(p)/L_b$ data set, presented in Figure 1b. The model reads, as follows,

$$\Delta L_b/L_b = -A_b\langle\varepsilon((T-T_c)/T_c, -(p-p_c(T))/p_c(T_c), u)\rangle_w + (\Delta L_b/L_b)_{off}. \quad (1)$$

Here, $A_b$ is the proportionality constant between the relative length change $\Delta L_b/L_b$ and the critical strain singlet $\varepsilon$ which represents the order parameter of an isostructural solid-solid transition. The strain singlet obeys the mean-field equation $r\varepsilon + u\varepsilon^3 = -\sigma$, with $r = (T-T_c)/T_c$ the reduced temperature, $\sigma = (p - p_c(T))/p_c$ the reduced pressure and $u$ a parameter, which quantifies non-linearities in the mean-field potential, and $\Delta L_b/L_b$ an offset contribution, see below. Note that the first two parameters are determined by our experiment, whereas the latter one enters as a free parameter. This model can be extended to account for the disorder broadening. To this end, we average the mean-field solution $\varepsilon(r,\sigma,u)$ with a Gaussian stress distribution $P_w(s) = 1/\sqrt{2\pi w}\exp\left(-s^2/(2w)\right)$ with variance $w$, i.e., $\langle\varepsilon(r,\sigma,u)\rangle_w = \int ds P_w(s)\varepsilon(r,\sigma+s,u)$. The details of this model are given in the Supplementary Information of Ref. [25]. In order to eliminate the temperature-dependent expansion effects, we normalized all data sets (see Figure 1) to the inflection points of $\Delta L_b(p)/L_b$ at $p_{\text{in}}$, i.e., $\Delta L_b(p_{\text{in}})/L_b = 0$. The increase in the broadening of the first-order transition upon irradiation is accompanied by an increasing asymmetry of the $\Delta L_b/L_b(p, T = \text{const.})$ curves in Figure 1b with respect to the midpoint of the discontinuity. To account for this effect, which is not understood at present and which is also not covered by the above model, we proceed as follows: we use the experimentally determined inflection point of the $\Delta L_b(p)/L_b$ data for a first iteration for $p_c(T)$ in the fitting procedure, and then refined the fit by allowing for a small variation of $p_c(T)$ of the order of 0.1 MPa. The resulting optimized $p_c(T)$ values are included as a brown broken line in the phase diagram in Figure 3. As a consequence of the small adjustment of $p_c(T)$, we have to accept a small offset in $\Delta L_b/L_b$ of $(\Delta L_b/L_b)_{off} = \mathcal{O}(10^{-6})$. The variance $w$ of the stress distribution can be determined from the experiment at lowest temperature of $T = 30$ K from a Gaussian fit to the $\kappa_b(p)$ data, which results in $w = 2.5 \times 10^{-4}$ for the data set of the irradiated sample. Note that this value of $w$ in its dimensionless form corresponds to the above-determined broadening of $\Delta P_{50h} = 0.81$ MPa.

Figure 4 demonstrates that the data set of the irradiated crystal is in very good agreement with the mean-field critical model, given in Equation (1), in the full $T$ range 30 K $\leq T \leq$ 40 K investigated. As a result of the fit, we obtain values for the parameters $u$ and $A_b$. The fit yields $u = 0.18$, which is the parameter characterizing the non-linearity of the length change. We stress that the fit to the data set on the pristine sample [25], which is included in Figure 4 for comparison, yielded the identical value of $u$. The extracted values $A_b$ are slightly different for the various $T$ = const. scans. However, they follow to a good approximation a $T$-linear behavior described by $A_b = (6.5 - 9.3(T-T_c)/T_c)) \times 10^{-5}$. The fits for the data set of the pristine sample resulted in a similar small $T$-linear variation of $A_b$. In the model for an isostructural solid-solid endpoint, the value of $A_b$ is given by the eigenvalues and the eigenvectors of the elastic constant matrix (see Ref. [25] for details) which is likely subject to a small $T$-dependent variation.

As demonstrated in Figure 4, the mean-field fits provide an excellent description of the experimental data not only for the pristine sample (see Figure 4), but also for the irradiated sample. These results show that the character of the Mott transition remains essentially unchanged for the two different disorder levels. This relates to the first-order character of the transition and the strong coupling to the lattice with the concomitant change of universality class to mean-field criticality around the second-order critical endpoint.

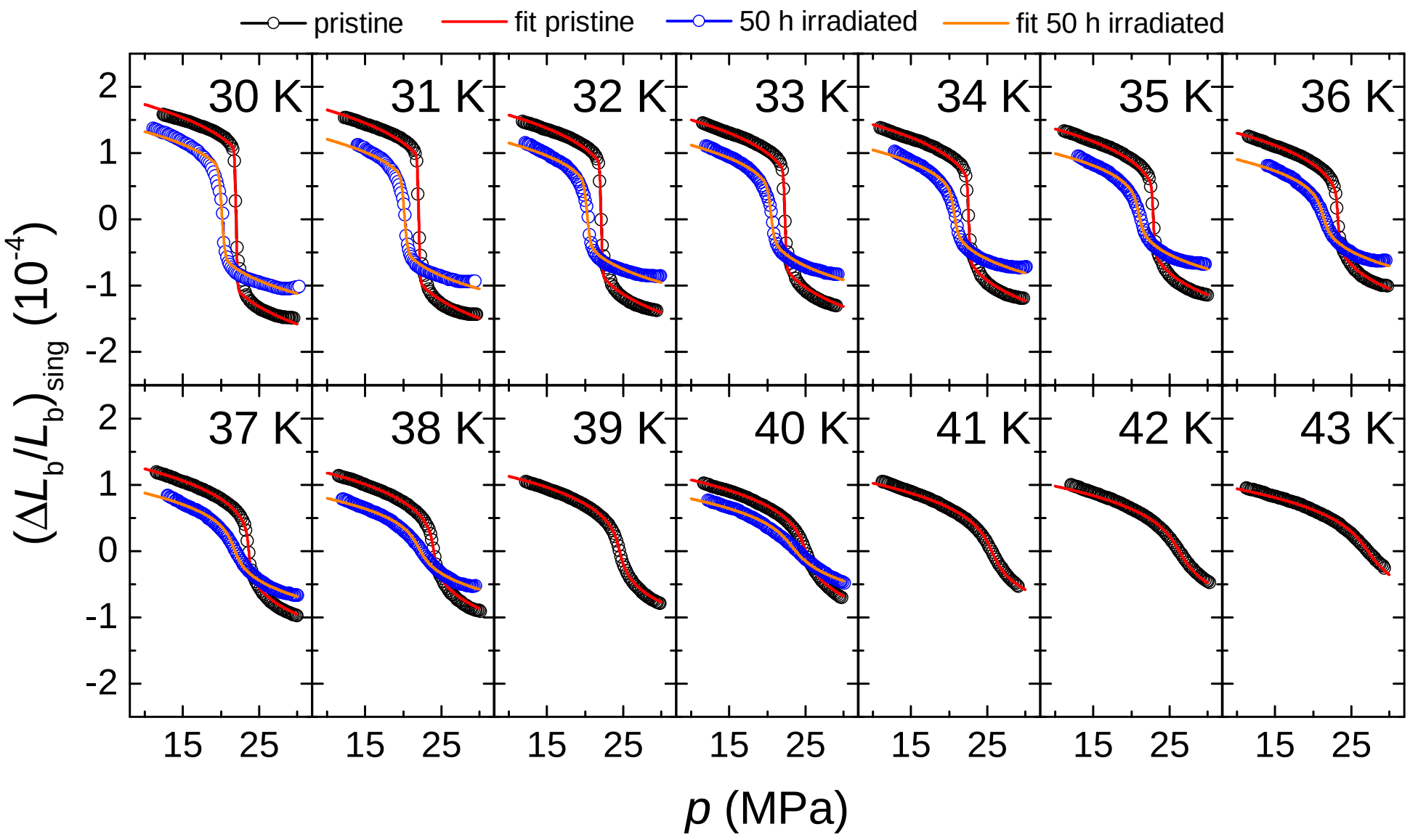

**Figure 4.** Singular part of the relative length change of $\kappa$-(BEDT-TTF)$_2$Cu[N(CN)$_2$]Cl along the out-of-plane $b$ axis, $(\Delta L_b/L_b)_{sing}$, (open symbols) as a function of pressure $p$ at various temperatures 30 K $\leq T \leq$ 43 K, together with a fit (straight lines) based on the mean-field solution, given by Equation (1). Black open symbols and red straight lines represent results on a pristine sample of batch #AF063, which were published in Ref. [25]. Blue open symbols and orange lines represent results on a crystal of batch #AF063 which was exposed to X-ray for 50 h.

### 3.4. *T-p Phase Diagrams of the Metal-Insulator Transition for Higher Irradiation Doses Based on Resistance Measurements*

In the following, we discuss how the characteristic features of the Mott transition for $\kappa$-Cl evolve upon further increasing the level of disorder. Instead of using measurements of the relative length change on one single crystal at varying irradiation levels, which is a very time consuming experiment, we employ measurements of the resistance $R$. Importantly, all of the presented data were taken on the same $\kappa$-Cl single crystal from batch #5-7 which was exposed to a three-step irradiation process (50 h, 100 h, and 150 h) and which was measured after each step.

Figure 5 shows the out-of-plane resistance $R$ as a function of $T$ at varying $p$=const. for a $\kappa$-Cl crystal from batch #5-7 after exposure to X-ray for 50 h (a), for 100 h (b) and 150 h (c). For all three irradiation levels, $\kappa$-Cl exhibits a non-metallic behavior at ambient pressure characterized by an increasing $R$ with decreasing $T$, consistent with literature results [48]. The $R(T, p = \text{const.})$ curves at different irradiation doses show the same general trend upon increasing $p$ as observed for pristine samples [13,14]: Low $p$ reduces the resistance at a fixed temperature, however, the resistance at a fixed, low $p$ value still shows a non-metallic behavior with $\mathrm{d}R/\mathrm{d}T < 0$. At intermediate $p$, the system is non-metallic at low $T$ and undergoes a transition into a metallic state, characterized by $\mathrm{d}R/\mathrm{d}T > 0$, upon increasing $T$ and enters an anomalous transport regime at higher $T$, characterized by $\mathrm{d}R/\mathrm{d}T \approx 0$. This reentrant behavior, visible in the range 18 MPa $\leq p \leq$ 30 MPa after 50 h irradiation time, reflects the $S$-shape of the first-order transition line and the *Widom* line in the $T$-$p$ phase diagram. At high $p >$ 30 MPa, all data sets reveal a metallic behavior in the full $T$ range investigated with a transition into a superconducting state at low $T \approx 12$ K.

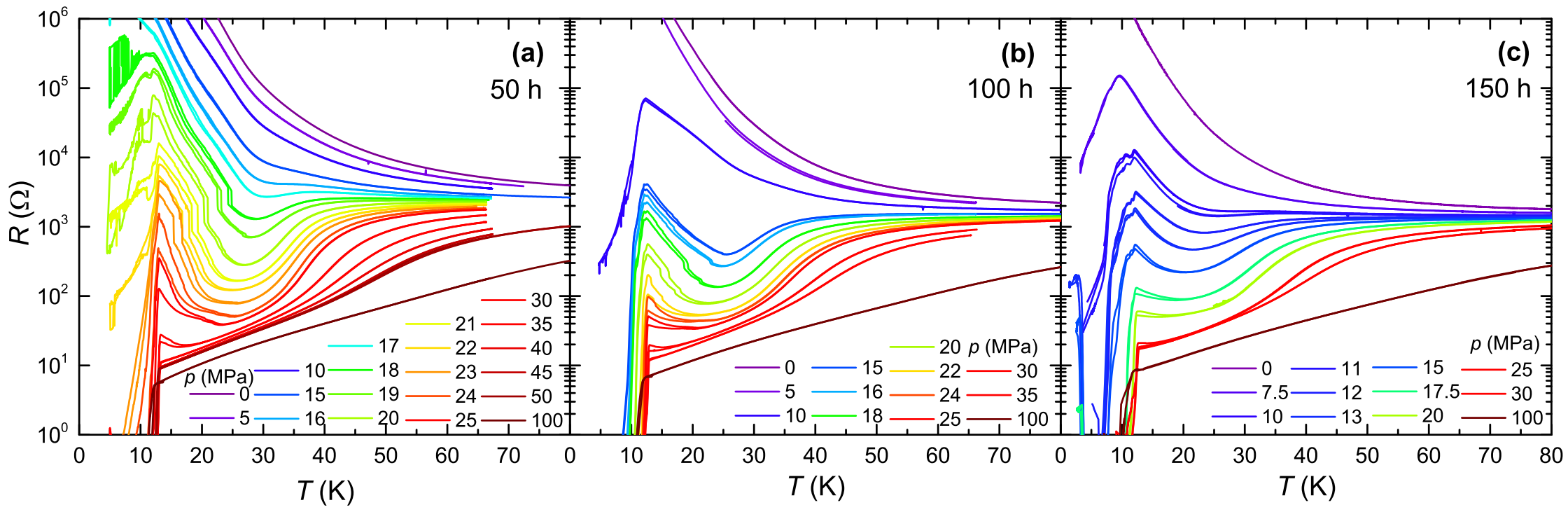

**Figure 5.** Out-of-plane resistance $R$ of $\kappa$-(BEDT-TTF)$_2$Cu[N(CN)$_2$]Cl (crystal #5-7), which was exposed to X-ray irradiation for 50 h (**a**); 100 h (**b**) and 150 h (**c**), as a function of temperature $T$ at various constant pressure values 0 MPa $\leq p \leq$ 100 MPa upon warming and upon cooling.

Despite these similarities in the overall behavior of $R(T, p = \text{const.})$ after different irradiation times, a closer inspection of the data in Figure 5 discloses significant differences in the location and character of the metal-insulator transition. These irradiation-induced changes can be illustrated clearly by comparing data sets taken at the same $p$ values, e.g., $p$ = 20 MPa. In the data at 50 h irradiation time, we find a kink in the $R(T, p = 20\,\text{MPa})$ data with an abrupt sign change of d$R$/d$T$ at $T \approx 12$ K which interrupts the increase in $R$ upon cooling, reflecting insulating behavior. This kink marks the entrance into a spurious superconducting phase at $T < 12$ K which is known to exist close to the first-order Mott metal-insulator boundary on the insulating side in pristine $\kappa$-Cl [11] and is likely responsible for the increased noise in $R(T,p = 20\,\text{MPa})$ at low $T$. Upon increasing the temperature, we find a pronounced jump-like suppression of $R(T)$ at $T \approx 20.5$ K. Importantly, the jump in $R(T, p = 20\,\text{MPa})$ is accompanied by a significant thermal hysteresis of $\Delta T \approx 0.7$ K, reflecting the first-order nature of this phase transition. We assign this discontinuity in $R(T, p)$ to the signature of the first-order metal-insulator transition, in accordance with the approach taken in literature [13,18,22]. The fact that the resistance jump remains sharp in the presence of some disorder is a consequence of percolation, i.e., a current that runs through low-resistance domains. This contrasts with the broadening observed in the discontinuities of thermodynamic quantities, such as the lattice effects, which average over all domains in the crystal. Even though the jump in $R(T)$ occurs at $T \approx 20.5$ K, we stress that the hysteresis and the insulating character are visible up to a slightly higher temperature of $T \approx 26.5$ K where $R(T, p = 20\,\text{MPa})$ exhibits a local minimum. Above the temperature of the local minimum in $R(T,\ p = 20\,\text{MPa})$, i.e., $T \approx 26.5$ K, the system shows metallic behavior with d$R$/d$T > 0$. Upon further increasing the temperature, the system gradually changes from a metal into an anomalous transport regime, as indicated by an almost flat $R(T, p = 20\,\text{MPa})$ behavior at $T \approx 38$ K. This change is not accompanied by any hysteresis, indicating that this temperature corresponds to a crossover temperature.

At a higher irradiation time of 100 h, cf. Figure 5b, the spurious superconducting phase at low $T$ and $p$ = 20 MPa is replaced by percolative or even bulk superconductivity with $T_{c,sc} \approx 12.6$ K, characterized by $R(T) < 10^{-2}\,\Omega$ at $T \approx 10.5$ K. With increasing temperature, we observe the same characteristics as in the case of 50 h irradiation, i.e., a jump-like reduction of the resistance at low $T$, accompanied by thermal hysteresis, and thus indicative of a first-order phase transition, and a crossover at high $T$. However, the jump in $R(T, p = 20\,\text{MPa})$ is significantly reduced in size and occurs in two concomitant steps at $T \approx 15.5$ K and $T \approx 17.5$ K. The fact that the transition takes place in several steps is likely of extrinsic nature and possibly associated with the formation of several domains in the presence of disorder. Therefore, we assign the metal-insulator transition to the larger jump at $T \approx 15.5$ K. The minimum in $R(T)$, which marks the onset of hysteresis, and the change from low-$T$ insulating to high-$T$ metallic behavior, is observed at $T \approx 22$ K. The crossover from the metallic behavior to the $R(T, p = 20\,\text{MPa}) \approx$ const. behavior is found at $T \approx 43$ K. Apparently, these characteristic temperatures react differently on increasing the irradiation dose from 50 h to 100 h: Whereas the position of the jump and the minimum in $R(T, p = 20\,\mathrm{MPa})$ are shifted to lower temperatures, the position of the crossover is shifted to higher $T$. This peculiar behavior can be related to the $S$-shaped form of the first-order transition line and the crossover lines (see e.g., Ref. [23,25]). In pristine $\kappa$-Cl, the temperature of the first-order transition decreases with increasing pressure distance to the critical point, whereas the crossover temperature increases. Thus, the present observation of a decrease in the first-order transition temperature and an increase in the crossover temperature at the same $p$ value upon irradiation can be considered as an indication that the critical endpoint of the irradiated sample is located lower in pressure than the pristine one.

Upon further increasing the irradiation time to 150 h, (cf. Figure 5c), we do not find any indications for a jump-like change of $R(T, p = 20\,\mathrm{MPa})$ at intermediate temperatures. Instead, the resistance changes smoothly yielding a minimum at $T = 16\,\mathrm{K}$. We note that despite the disappearance of a jump-like change of $R(T, p = 20\,\mathrm{MPa})$, we still find a small hysteresis at low $T$ in the insulating regime which prevails until the resistance reaches its minimum. Apparently, at this irradiation level, the character of the phase transition has changed. Based on the present transport data, we cannot discriminate between a transition of weak first order, or a smeared first-order transition with residual hysteresis.

In order to illustrate the change of the first-order metal-insulator transition line upon irradiation, we compile the position of the jump for all investigated $p$ values in a $T$-$p$ phase diagram in Figure 6. For comparison, we included the transition temperatures of the first-order transition line extracted from $T$-dependent resistance measurements on a pristine crystal of batch #AF063 (not shown) which are in good agreement with published data [11,14,22,25]. As already indicated by our thermodynamic measurements, we find a significant shift of the first-order transition line to lower $p$ upon irradiation for 50 h and 100 h. The shift of the phase transition line at $T = 20\,\mathrm{K}$, amounts to about $-(0.07 \pm 0.01)\,\mathrm{MPa}$ per hour of irradiation time. In addition, we find that the position of the critical endpoint ($T_c$,$p_c$) becomes significantly reduced in temperature as well as in pressure: Our analysis yields $(T_c,p_c) \approx ((29 \pm 1)\,\mathrm{K}, (17.5 \pm 0.5)\,\mathrm{MPa})$ and $(T_c,p_c) \approx ((27 \pm 1)\,\mathrm{K}, (15.0 \pm 0.5)\,\mathrm{MPa})$ after irradiating $\kappa$-Cl for 50 h and 100 h, respectively. We stress that the slightly different results of ($T_c$,$p_c$) after 50 h of irradiation, inferred from the resistance measurements on a crystal of batch #5-7, as compared to those of our thermodynamic measurements on a crystal of batch #AF063 (see Section 3.2), might be related to details of our experiments. First, we cannot rule out small sample-to-sample variations of crystals from different batches. More importantly, however, the crystals involved had different surface-to-thickness ratios, implying that a different distribution of disorder was introduced within the same irradiation time. At even higher irradiation time of 150 h, we cannot find any discontinuous feature in the full $T$-$p$ range investigated. Therefore, we cannot draw a phase transition line for this case.

To quantify the range of hysteretic behavior and to follow its evolution upon irradiation, we define the onset of hysteresis by $\frac{R_{warm}(T)-R_{cool}(T)}{R_{warm}(T)} \geq 0.025$ with $R_{warm}$ ($R_{cool}$) denoting the resistance taken upon warming (cooling). The onset temperatures together with the previously determined first-order transition lines (green symbols) and the second-order critical endpoint (thick green circle) are compiled in the contour plots in Figure 7 which will be discussed below in more detail. For the 50 h irradiated crystal, we find clear signatures for thermal hysteresis in the pressure range $\approx 15\,\mathrm{MPa} \leq p \leq \approx 30\,\mathrm{MPa}$ and up to $T \approx 29\,\mathrm{K}$. The width of the hysteresis range (hatched area in Figure 7) on the pressure axis is similar to the one found for a pristine crystal in Ref. [13]. However, upon irradiation, the hysteresis regime is shifted to lower pressures and suppressed in temperature, consistent with the shift of the Mott transition line to lower $p$ and the reduction of $(T_c, p_c)$, discussed above. Upon further increasing the irradiation to 100 h, the hysteresis regime is shifted to even lower pressures and further reduced in temperature. Importantly, even after 150 h of irradiation thermal hysteresis occurs in a similarly wide pressure range $\approx 10\,\mathrm{MPa} \leq p \leq \approx 30\,\mathrm{MPa}$ and up to $T \approx 20\,\mathrm{K}$. This result speaks in favor of the notion of a smeared first-order transition after 150 h of irradiation where disorder causes the suppression of signatures, and therefore can no longer be discerned from a crossover. We stress that for a detailed analysis of the hysteresis range, measurements as a function of increasing and decreasing pressure, not possible with the available set-up, are much better suited for the following reasons: First, these organic charge-transfer salts tend to form small cracks due to their large thermal expansion, particularly when sweeping the temperature over a wide range, as done in the present work. These small cracks lead to small offset contributions in $R(T, p)$ which strongly hamper the analysis of the thermal hysteresis. This effect is avoided in pressure sweeps performed at low temperatures. Second, due to the almost vertical alignment of the metal-insulator transition line in the $T$-$p$ phase diagram, the signatures in $p$ sweeps are more clearly pronounced as compared to $T$ sweeps.

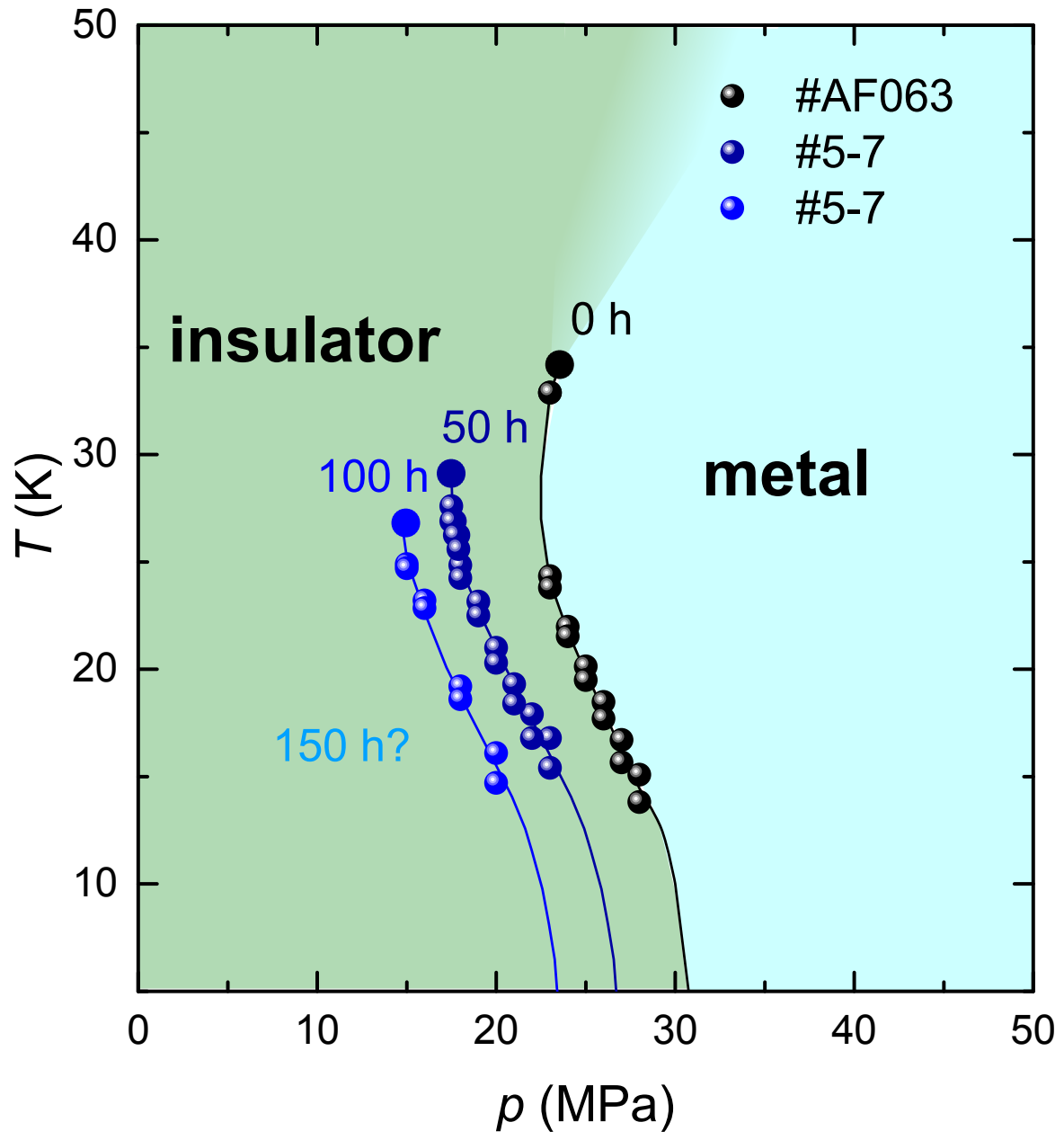

**Figure 6.** $T$-$p$ phase diagram of the first-order Mott metal-insulator transition line of $\kappa$-(BEDT-TTF)$_2$Cu[N(CN)$_2$]Cl upon increasing disorder which was intentionally introduced by X-ray irradiation. Black circles correspond to the phase transition line obtained from measurements (not shown) on a pristine sample of batch #AF063. Dark blue and light blue circles correspond to the phase transition line obtained from measurements on the same sample of batch #5-7 after exposure to X-ray for 50 h and 100 h, respectively (see Figure 5). Note that pairs of data points taken at the same $p$ value correspond to the position of the jump upon warming and cooling. The lines are guides to the eyes. In case of 150 h no discontinuities in $R(T, p)$ could be resolved signaling the absence of a strong first-order phase transition.

### *3.5. Influence of Disorder on the Metallic and Insulating States Nearby the Mott Transition*

Next, we focus on the influence of disorder on the range close to the first-order Mott metal-insulator transition. To this end, we include in Figure 7 contour plots of the resistance in a wide $T$-$p$ range for all three investigated irradiation times. The plots also contain the transition temperatures into spurious (dark grey symbols) or bulk (light grey symbols) superconductivity. The colors represent the absolute value of $R$ on a logarithmic scale. This presentation of the data highlights two remarkable aspects. First, the diagram shows how the superconducting state, observed in the immediate vicinity of the metal-insulator transition in the pristine case, is affected by irradiation. Second, it discloses a striking mirror symmetry along the *Widom* line (see below for detailed definition) for $T \gg T_c$ which has been known from the pristine case [29]. In the latter case, this mirror symmetry has been interpreted in terms of a quantum-critical scaling [28].

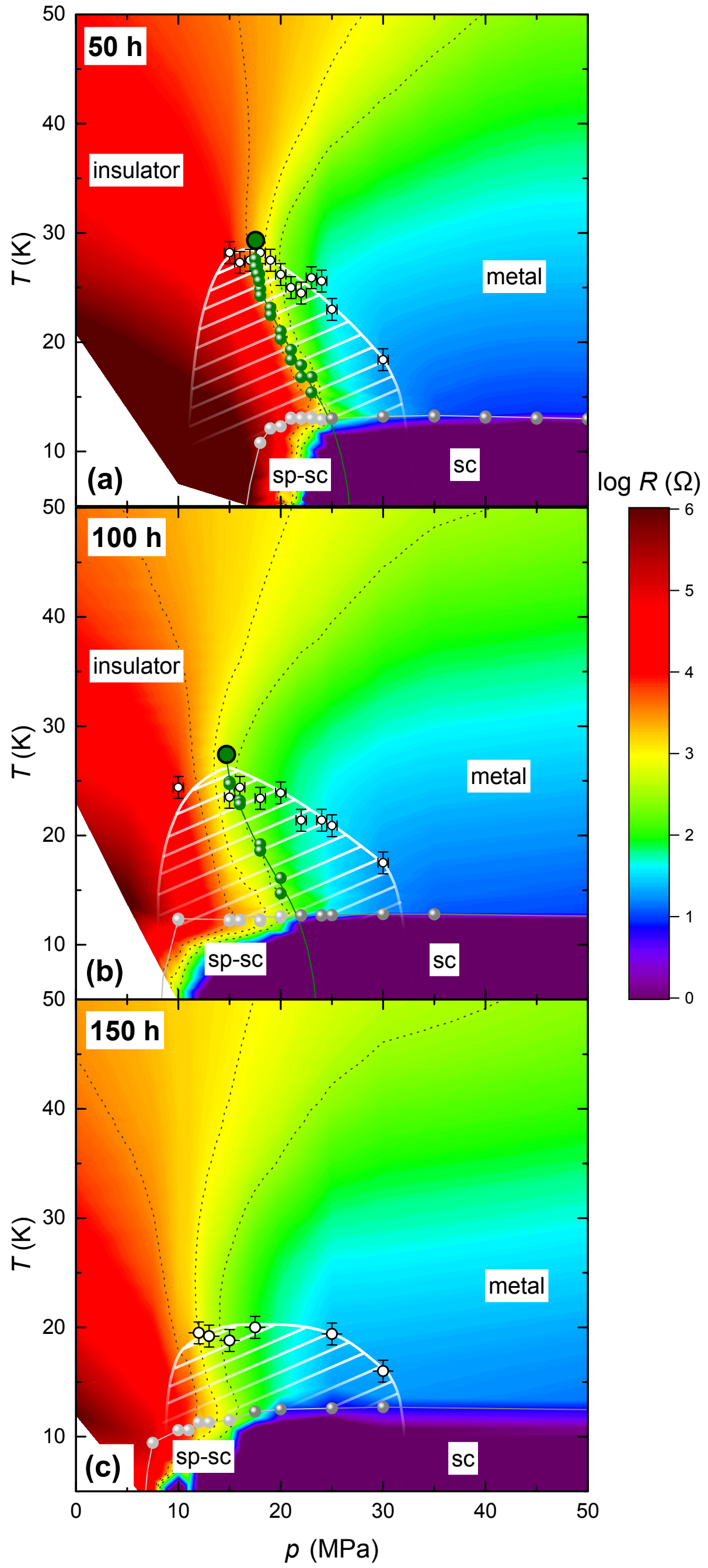

**Figure 7.** Experimentally determined $T$-$p$ phase diagrams for a $\kappa$-(BEDT-TTF)$_2$Cu[N(CN)$_2$]Cl crystal of batch #5-7 after exposure to X-ray for 50 h (**a**); 100 h (**b**) and 150 h (**c**). Data were extracted from the resistance $R$ measurements presented in Figure 5. Green symbols in (**a**,**b**) represent the first-order transition line extracted from cooling as well as warming experiments, the black-edged big green circle marks the second-order critical endpoint. After exposure to X-ray for 150 h (**c**) no discontinuities in $R(T,p)$ indicating a first-order phase transition could be resolved. White circles indicate the temperature, below which a sizable hysteresis between warming and cooling was observed. White hatched area is a guide to the eye for the hysteresis region, delimited by the white circles. Dark grey symbols mark the entrance into a spurious superconducting (sp-sc) state. Light grey symbols indicate the transition into a bulk superconducting (sc) state. Lines are guide to the eyes. The background shows a contour plot of $\log(R(T,p)/\Omega)$. Dashed lines indicate three constant-resistance lines with $\log(R/\Omega) = 2.5\,\Omega$, $\log(R/\Omega) = 3\,\Omega$ and $\log(R/\Omega) = 3.5\,\Omega$.

Concerning the superconducting properties of irradiated $\kappa$-Cl, we find signatures for a bulk superconducting dome as well as a range of spurious superconductivity in the $T$-$p$ phase diagram for all three irradiation doses. We use the term bulk superconductivity in cases where there is a sharp drop of $R$ at $T_{c,sc}$ below which the out-of-plane resistance reaches values $R(T,p) \ll 1\,\Omega$ for $T < T_{c,sc}$. This contrasts with spurious superconductivity which manifests itself in a broad kink in $R(T,p)$ at low $T$ without reaching a low resistance value, i.e., $R(T,p) > 10\,\Omega$. After 50 h (100 h) of irradiation, we find spurious superconductivity in the range 18 MPa $\leq p \leq$ 24 MPa (10 MPa $\leq p \leq$ 20 MPa) and bulk superconductivity in the range $p \geq$ 25 MPa ($p \geq$ 22 MPa). Compared to the range of spurious superconductivity of 20 MPa $\leq p \leq$ 28 MPa and bulk superconductivity of $p \geq$ 30 MPa in pristine $\kappa$-Cl [11], we find a significant shift of superconductivity to lower pressures upon increasing irradiation dose. We note that the shift of the superconducting dome on the pressure axis is similar to the shift of the first-order Mott metal-insulator transition line: The spurious superconducting phase occurs in all cases on the insulating side of the Mott transition, whereas the bulk superconducting phase can be found right on the metallic side. The fact that bulk superconductivity emerges as soon as the insulator is suppressed by pressure indicates the strong connection between the metal-insulator transition and superconductivity, also in disordered $\kappa$-Cl. Surprisingly, even after 150 h of irradiation, bulk superconductivity is observed in a wide $p$ range despite the lack of clear signatures of a first-order phase transition. However, superconductivity starts to occur in the regime where the resistance at $T > T_{c,sc}$ changes strongly in a narrow pressure regime. This result suggests that the existence of superconductivity does not require a nearby strong first-order metal-insulator transition.

Besides the irradiation-induced shift of the superconducting dome on the pressure axis, the critical temperature $T_{c,sc}$ is also affected by irradiation. By using the onset temperature of the drop in $R(T)$ as a measure of $T_{c,sc}$, we find that the maximum $T_{c,sc}$ is reduced from 13 K after 50 h of irradiation, to 12.7 K after 100 h and 12.2 K after 150 h. At $p = 100$ MPa, the highest pressure of the present experiment, $T_{c,sc}$ is reduced from 12.3 K to 12.0 K to 11.8 K upon increasing irradiation time from 50 h to 100 h to 150 h. A similar reduction of $T_{c,sc}$ upon irradiation was found also for $\kappa$-(BEDT-TTF)$_2$Cu(NCS)$_2$ [49,51], and $\kappa$-(BEDT-TTF)$_2$Cu[N(CN)$_2$]Br [50], both of which are located at ambient pressure on the metallic side of the Mott metal-insulator transition. In the latter cases, this shift was interpreted to be incompatible with a superconducting order parameter with one single component. This conclusion is consistent with the result of a recent combined theoretical and experimental scanning tunneling microscopy study [57] which found strong indications for an extended $s + d_{x^2-y^2}$ symmetry of the superconducting order parameter in $\kappa$-(BEDT-TTF)$_2$Cu[N(CN)$_2$]Br.

Now we turn to the analysis of resistance behavior in the high-$T$ regime $T \gg T_c$. To this end, we included in the contour plots in Figure 7 three dotted lines which represent constant $R$ lines in the phase diagram with $\log(R/\Omega) = 3.5$, $\log(R/\Omega) = 3$ and $\log(R/\Omega) = 2.5$. The $\log(R/\Omega) = 3$ is chosen such that it is very close to the resistance value along the *Widom* line which is determined according to Refs. [28,29] as the inflection points of $\log R$ vs. $p$ at $T = \text{const.}$ For all three irradiation doses, we find an extended range at $T > T_c$ where $\log R$ shows a mirror symmetry along the *Widom* line. The notation of mirror symmetry here implies that the normalized resistance $\log R_{norm}(\delta p, T = \text{const.}) = |\log R(p, T = \text{const.}) - \log R(p_{widom}, T = \text{const.})| / \log R(p_{widom}, T = \text{const.})$ with $\delta p = p - p_{widom}$, i.e., the resistance normalized to the value at the *Widom* line at $p_{widom}$, are identical on the low-pressure side of the *Widom* line ($\delta p < 0$) and the high-pressure side ($\delta p > 0$) for a given pressure distance $\delta p$ to the *Widom* line ($\log R_{norm}(-\delta p) = \log R_{norm}(\delta p)$). The temperature range of this symmetric $\log R(T,p)$ behavior is extended to lower $T$ upon increasing irradiation: After 150 h of irradiation this symmetry is visible down to $T \approx 20$ K. As this mirror symmetry was associated with signatures of quantum criticality for pristine $\kappa$-Cl [29], the present data indicate that irradiation can be an appropriate parameter to further suppress $T_c$, thereby expanding the quantum-critical regime to somewhat lower temperatures. However, even after 150 h of irradiation, it appears that $T_c$ is not fully suppressed to zero (see discussion of hysteresis above), so that more disorder is required to fully suppress the second-order critical endpoint in $\kappa$-Cl. In this situation, it is possible that quantum-critical signatures of the Mott transition can develop. However, it may well be that in this case, the material's properties are governed by disorder effects rather than quantum criticality. For a detailed investigation of the potential quantum-critical scaling around the *Widom* line and of the role of disorder on this scaling (see Ref. [58]), in-plane resistance measurements up to higher $T$, as well as measurements of $R$ as a function of $p$, similar to Ref. [29], are required.

## 4. Discussion

In the following, we discuss our experimental observations on the metal-insulator transition for irradiated $\kappa$-Cl in the context of theoretical results on the Mott-Anderson transition, obtained from combining dynamic mean-field theory with different numerical implementations of disorder [36,37,41,59].

One of the central results of the present work is the observation that the character of the Mott metal-insulator transition remains essentially unchanged when disorder is introduced to a moderate extent, corresponding to an irradiation time of about 100 h. This relates to the first-order character of the phase transition, as revealed by thermal expansion and resistance measurements, and the second-order critical endpoint of this line at ($T_c$,$p_c$). We showed that after 50 h of irradiation, strong lattice effects dominate the critical behavior around ($T_c$, $p_c$), similar to the pristine case [25]. Thus, we conclude that exposure to X-ray for 50 h and 100 h induces a degree of disorder which is small enough to conserve essential Mott properties. This interpretation is supported by theoretical results [37,41,59] which indicate a sharp metal-insulator transition, accompanied by a sizable coexistence regime, at a critical interaction strength $U_c$ at $T = 0$, similar to the clean Mott system.

However, even though retaining its first-order character, we find that the transition becomes distinctly weaker upon irradiation, as evidenced by a significant shrinkage of the discontinuities in the length change and resistance in the 50 h and 100 h irradiated samples, as compared to the pristine ones. Moreover, after 150 h of irradiation, no discontinuity could be resolved in $R(T, P)$ anymore, implying a significant change of the character of the Mott transition. These observations are in line with theoretical predictions: The calculations reveal a sizable reduction of the size of the coexistence region upon an initial increase in disorder [37], reflecting the weakening of the first-order transition. In addition, recent finite-temperature calculations suggest a sudden drop of $T_c$ to zero [41] and a continuous change from a metal to an insulator above some critical value of the degree of disorder. The present data do not allow to conclude that the character of the transition changes from first to second order. However, it seems likely that a critical level of disorder, above which the discontinuous first-order character of the Mott transition is suppressed, is attained in $\kappa$-Cl between 100 h and 150 h of irradiation. To investigate the character of the phase transition above 100 h of irradiation, a detailed study of the size of the coexistence region for different disorder levels can be helpful. Such a study has to include pressure-dependent hysteresis measurements around the metal-insulator transition, as explained in detail above.

Furthermore, we find a continuous shift of the first-order Mott transition line to lower pressures on going from the pristine sample to the ones with 50 h and 100 h irradiation time. This decrease of $p_c$ corresponds to an increase of the critical correlation strength $(U/W)_c$. Theoretically, it has been predicted that disorder widens the Hubbard bands which causes an increase of spectral weight at the Fermi level [37]. As a result, a larger $U$ is required for opening the Mott gap, equivalent to the experimentally-observed increase in $(U/W)_c$. This situation, where disorder induces a finite spectral weight in the clean Mott gap, has been labeled as the *soft Coulomb gap* scenario [38,60]. Thus, the present results are consistent with the scenario that disorder changes the fully-gapped Mott insulator $\kappa$-Cl to a Mott insulator with a soft Coulomb gap. We note that a similar conclusion was drawn from the analysis of optical conductivity data on the same compound after irradiation [48] as well as from scanning tunneling microscopy studies of the normal-state density of states in $\kappa$-(BEDT-TTF)$_2$Cu[N(CN)$_2$]Br even in its pristine form [61]. At the same time, our finding points towards a possibility to estimate the degree of disorder for a given sample of $\kappa$-Cl by measuring the position of the first-order transition line in the $T$-$p$ phase diagram. However, as this approach is unable to provide an absolute measure of the degree of disorder, it allows only to compare the relative disorder level of different crystals. Concerning the effect of disorder on the critical temperature $T_c$ of the second-order critical endpoint of the Mott transition, we observe a significant suppression of $T_c$ by a few Kelvin upon increasing the irradiation up to 100 h. Also this trend is consistent with the results of finite-temperature calculations [41,59] on a strongly correlated electron system with weak disorder.

On a qualitative level, the experimental results agree well with all presently available theoretical predictions for the metal-insulator transitions within the Mott-Anderson model. For a more quantitative comparison, which is required to distinguish between the different theoretical calculations, however, a detailed characterization of the disorder is required. This includes information on the nature, the concentration and spatial distribution of the disorder [62]. According to infrared absorption measurements on $\kappa$-(BEDT-TTF)$_2$Cu[N(CN)$_2$]Br (see Figure 2 in Ref. [48]), exposure to X-ray with a dose rate of 0.5 MGy/h, as used in the present work (see Methods Section 2), destroys C–N bonds in the anion layer at a rate of ≈10–20 % per 50 h of irradiation. Another sensitive probe on the degree of disorder is the residual resistivity ratio ($RRR$). Our data of the resistivity reveal a decrease of the $RRR = R(65\,\mathrm{K})/R(12\,\mathrm{K})$ in the metallic phase at 100 MPa upon irradiation from $RRR \approx 41$ after 50 h irradiation to ≈24 after 100 h and ≈18 after 150 h. This observation is fully consistent with an increase of disorder upon irradiation. Open questions relate on the one hand to the penetration depth of the X-ray photons and the resulting spatial distribution of the introduced disorder. Besides that, a deeper understanding of the interaction between the disorder in the anion layers and the delocalized electrons in the BEDT-TTF layers seems necessary for a comparison with theoretical results which consider disorder embedded directly in the electronic system.

Finally, we would like to point the attention towards interesting future research directions in the field of correlated electrons in the presence of disorder. As mentioned above, the present data suggest the possibility to suppress the critical endpoint significantly after 150 h of irradiation suggesting the possibility to fully suppress $T_c$ to zero at even higher irradiation. In this situation, it would be of high interest to investigate how the quantum-critical signatures, observed in the pristine case only at higher $T$, evolve towards lower temperatures [58]. In addition, the interplay of charge- with spin- degrees of freedom in the limit of strong disorder and strong correlations [39], as represented by the data set taken after 150 h irradiation, hold promise for interesting new physics. This is demonstrated, e.g., in experiments on $\kappa$-Cl after 500 h of irradiation where long-range magnetic order is replaced by a state with spin-liquid properties [53]. In the present study, we did not investigate the magnetic properties of the irradiated samples. However, it has been known that the inflection point of the first-order transition line $T_{MI}(p)$ at $T \approx 27\,\mathrm{K}$ in the pristine case is connected to the occurrence of long-range magnetic ordering due to entropy reasons [63,64]. As demonstrated in Figure 6, the inflection point of $T_{MI}(p)$ shifts to lower temperatures upon increasing irradiation up to 100 h, indicating a lowering of the magnetic ordering temperature. This is consistent with previous reports of the decreased magnetic ordering temperature in irradiated samples of $\kappa$-Cl at ambient pressure [65]. To investigate this ordering at higher irradiation, where disorder causes a change of the character of the Mott transition, and its interplay with the metal-insulator transition, we suggest to study the magnetic properties of strongly irradiated $\kappa$-Cl under pressure.

## 5. Conclusions

In summary, we have studied the effect of X-ray-induced disorder on the Mott metal-insulator transition for the organic charge-transfer salt $\kappa$-(BEDT-TTF)$_2$Cu[N(CN)$_2$]Cl. The application of hydrostatic He-gas pressure allowed us to fine-tune the system across the Mott transition. Our main findings, based on thermal expansion and resistance measurements, include (i) strong lattice effects around the Mott transition after irradiating the system for 50 h; (ii) a line of first-order transitions which ends in a second-order critical endpoint after 50 h and 100 h irradiation with a significant lowering of the critical temperature $T_c$ and the critical pressure $p_c$ upon increasing irradiation and (iii) a change of the character of the Mott transition for 150 h irradiation where no discontinuous signatures of a first-order phase transition could be detected despite the presence of a small thermal hysteresis. The results (i) and (ii) reveal a very similar behavior of the irradiated samples to the pristine case, i.e., a strong coupling of the lattice to the electronic degrees of freedom close to the second-order critical endpoint. This observation speaks in favor of a dominant Mott character even after 50 h and 100 h of irradiation indicating that the Mott transition in $\kappa$-(BEDT-TTF)$_2$Cu[N(CN)$_2$]Cl is stable against the introduction of a small degree of disorder. The reduction in $p_c$ and $T_c$ after irradiation is fully compatible with theoretical predictions of the phase diagram for the Mott-Anderson transition. The latter indicate a larger critical $U$ to open the Mott gap due to a disorder-induced increase of spectral weight at the Fermi level, corresponding to a smaller $p_c$. Thus, our experimental results are fully consistent with the theoretically predicted *soft Coulomb gap* scenario for a strongly correlated electron system with small amount of disorder. In addition, result (iii) reflects the strong weakening of the first-order character of the metal-insulator transition upon increasing irradiation which eventually may turn the transition into a crossover between the insulator and the metal. For future investigations and a more quantitative comparison between theory and experiment, a detailed knowledge of the spatial distribution of X-ray-induced disorder in the anion layer and its interplay with the electrons in the BEDT-TTF layer is necessary.

Besides the investigation of the metal-insulator transition at low $T$, we were able to show that superconductivity close to the metal-insulator transition prevails for all investigated disorder levels. The superconducting dome shifts similarly to the first-order metal-insulator transition to lower $p$ upon irradiation. Future studies on the interplay of the charge with spin degrees of freedom, and the potential quantum-critical signatures when $T_c$ is further suppressed, hold promise for interesting new physics in the field of strongly correlated electrons in the presence of disorder.

**Acknowledgments:** Research in Frankfurt and Dresden was supported by the German Science Foundation via the Transregional Collaborative Center SFB/TR49 "Condensed Matter Systems with Variable Many-Body Interactions" and Collaborative Research Center SFB 1143 "Correlated Magnetism: From Frustration to Topology". Work in Sendai was partly supported by the Japan Society for the Promotion of Science KAKENHI grant #JP25287080.

**Author Contributions:** M.L. and T.S. conceived and supervised the project. E.G. performed thermal expansion under pressure experiments. U.T., A.N. and S.K. performed resistance under pressure experiments. M.G. provided theoretical support in analyzing the thermal expansion data. H.S. and T.S. grew the single crystals. T.S. performed the irradiation of the crystals. E.G. and M.L. wrote the paper with contributions from U.T., T.S. and M.G.

**Conflicts of Interest:** The authors declare no conflict of interest.